\documentclass[aps,prb,twocolumn,superscriptaddress,preprintnumbers,amsmath,amssymb]{revtex4-1}  
\usepackage{graphicx}  
\usepackage{dcolumn}   
\usepackage{bm}        
\usepackage{color}
\usepackage{epstopdf}
\usepackage{float}

\begin{document}


\title{Engineering the structural and electronic phases of MoTe$_2$ through W substitution}

\author{D. Rhodes$^{\dag}$}
\affiliation{National High Magnetic Field Laboratory, Florida State University, Tallahassee, FL 32310, USA}
\affiliation{Department of Physics, Florida State University, Tallahassee-FL 32306, USA}
\author{D. A. Chenet$^{\dag}$}
\affiliation{Department of Mechanical Engineering, Columbia University, New York, NY 10027, USA}
\author{B. E. Janicek}
\affiliation{Department of Materials Science and Engineering, University of Illinois Urbana−Champaign, Urbana, IL 61801, USA}
\author{C. Nyby}
\affiliation{Department of Chemistry, Stanford University, Stanford University, Stanford, CA 94305-4401, USA}
\author{Y. Lin}
\affiliation{Department of Applied Physics and Applied Mathematics, Columbia University, New York, NY 10027, USA}
\author{W. Jin}
\affiliation{Department of Applied Physics and Applied Mathematics, Columbia University, New York, NY 10027, USA}
\author{D. Edelberg}
\affiliation{Department of Physics, Columbia University, New York, NY 10027, USA}
\author{E. Mannebach}
\affiliation{Department of Materials Science and Engineering, Stanford University, Stanford, CA 94305, USA}
\author{N. Finney}
\affiliation{Department of Mechanical Engineering, Columbia University, New York, NY 10027, USA}
\author{A. Antony}
\affiliation{Department of Mechanical Engineering, Columbia University, New York, NY 10027, USA}
\author{T. Schiros}
\affiliation{Columbia University, Materials Research Science and Engineering Center, NY, NY 10027 USA}
\affiliation{SUNY Fashion Institute of Technology, Department of Science and Mathematics, NY, NY 10001 USA}
\author{T. Klarr}
\affiliation{Sensors and Electronic Devices Directorate  US Army Research Laboratory Adelphi, MD 20723, USA}
\author{A. Mazzoni}
\affiliation{Sensors and Electronic Devices Directorate  US Army Research Laboratory Adelphi, MD 20723, USA}
\author{M. Chin}
\affiliation{Sensors and Electronic Devices Directorate  US Army Research Laboratory Adelphi, MD 20723, USA}
\author{Y.-c Chiu}
\affiliation{National High Magnetic Field Laboratory, Florida State University, Tallahassee, FL 32310, USA}
\affiliation{Department of Physics and National High Magnetic Field Laboratory, Florida State University, Tallahassee, FL 32310, USA}
\author{W. Zheng}
\affiliation{National High Magnetic Field Laboratory, Florida State University, Tallahassee-FL 32310, USA}
\affiliation{Department of Physics and National High Magnetic Field Laboratory, Florida State University, Tallahassee, FL 32310, USA}
\author{Q. R. Zhang}
\affiliation{National High Magnetic Field Laboratory, Florida State University, Tallahassee-FL 32310, USA}
\affiliation{Department of Physics and National High Magnetic Field Laboratory, Florida State University, Tallahassee, FL 32310, USA}
\author{F. Ernst}
\affiliation{Department of Applied Physics, Stanford University, Stanford, CA 94305-4090, USA}
\affiliation{Stanford PULSE Institute, SLAC National Accelerator Laboratory, Menlo Park, CA 94025, USA}
\author{J. I. Dadap}
\affiliation{Department of Electrical Engineering, Columbia University, New York, NY 10027, USA}
\author{X. Tong}
\affiliation{Center for Functional Nanomaterials, Brookhaven National Laboratory, Upton, NY 11973-5000, USA}
\author{J. Ma}
\affiliation{Beijing National Laboratory for Condensed Matter Physics, and Institute of Physics, Chinese Academy of Sciences, Beijing 100190, China}
\author{R. Lou}
\affiliation{Department of Physics, Renmin University of China, Beijing 100872, China}
\author{S. Wang}
\affiliation{Department of Physics, Renmin University of China, Beijing 100872, China}
\author{T. Qian}
\affiliation{Beijing National Laboratory for Condensed Matter Physics, and Institute of Physics, Chinese Academy of Sciences, Beijing 100190, China}
\author{H. Ding}
\affiliation{Beijing National Laboratory for Condensed Matter Physics, and Institute of Physics, Chinese Academy of Sciences, Beijing 100190, China}
\author{R. M. Osgood, Jr}
\affiliation{Department of Electrical Engineering, Columbia University, New York, NY 10027, USA}
\affiliation{Department of Applied Physics and Applied Mathematics, Columbia University, New York, NY 10027, USA}
\author{D. W. Paley}
\affiliation{Department of Chemistry, Columbia University, New York, NY 10027, USA}
\affiliation{Columbia Nano Initiative, Columbia University, New York, NY 10027, USA}
\author{A. M. Lindenberg}
\affiliation{Department of Materials Science and Engineering, Stanford University, Stanford, CA 94305, USA}
\affiliation{Stanford PULSE Institute, SLAC National Accelerator Laboratory, Menlo Park, CA 94025, USA}
\affiliation{Stanford Institute for Materials and Energy Sciences, SLAC National Accelerator Laboratory, Menlo Park, CA 94025, USA}
\author{P. Y. Huang}
\affiliation{Department of Mechanical Science and Engineering, University of Illinois Urbana−Champaign, Urbana, IL 61801, USA}
\author{A. N. Pasupathy}
\affiliation{Department of Physics, Columbia University, New York, NY 10027, USA}
\author{M. Dubey}
\affiliation{Sensors and Electronic Devices Directorate  US Army Research Laboratory Adelphi, MD 20723, USA}
\author{J. Hone}
\affiliation{Department of Mechanical Engineering, Columbia University, New York, NY 10027, USA}
\author{L. Balicas}
\email[]{balicas@magnet.fsu.edu}
\affiliation{National High Magnetic Field Laboratory, Florida State University, Tallahassee, FL 32310, USA}



\begin{abstract}
\textbf{MoTe$_2$ is an exfoliable transition metal dichalcogenide (TMD) which crystallizes in three symmetries;
the semiconducting trigonal-prismatic $2H-$phase, the semimetallic $1T^{\prime}$ monoclinic phase, and the semimetallic
orthorhombic $Td$ structure \cite{Revolins,Vellinga,review1,MoTe2_MI_transition}. The $2H-$phase displays a band gap of $\sim 1$ eV\cite{Heinz}
making it appealing for flexible and transparent optoelectronics. The $Td-$phase is predicted to possess unique topological properties
\cite{bernevig,felser,bernevig2,TP_transition} which might lead to topologically protected
non-dissipative transport channels \cite{TP_transition}. Recently, it was argued that it is possible to locally induce
phase-transformations in TMDs \cite{review1,reed1,reed2,Reed}, through chemical doping \cite{phase_engineering}, local heating \cite{phase_patterning}, or electric-field \cite{Reed,zhang} to achieve ohmic contacts or to induce useful functionalities such as electronic
phase-change memory elements\cite{reed2}.  The combination of semiconducting and topological elements based upon the same compound,
might produce a new generation of high performance, low dissipation optoelectronic elements.
Here, we show that it is possible to engineer the phases of MoTe$_2$ through W substitution by unveiling the
phase-diagram of the Mo$_{1-x}$W$_x$Te$_2$ solid solution which displays a semiconducting to semimetallic transition as a function of $x$.
We find that only $\sim 8$ \% of W stabilizes the $Td-$phase at room temperature. Photoemission spectroscopy, indicates that this phase possesses a Fermi surface
akin to that of WTe$_2$ \cite{pletikosic}.}
\end{abstract}
\maketitle

The properties of semiconducting and of semimetallic MoTe$_2$ are of fundamental interest in their own right, but are also for their potential technological relevance. In the mono- or few-layer limit it is a direct-gap semiconductor, while the bulk has an indirect bandgap
\cite{Heinz,Lezama,Morpurgo_crossover} of $\sim$ 1 eV. The size of the gap is similar to that of Si, making $2H-$MoTe$_2$ particularly appealing for both purely electronic devices \cite{MoTe2_FETs,MoTe2_ambipolar}  and optoelectronic applications \cite{Mak_review}. Moreover, the existence of different phases opens up the possibility for many novel devices and architectures. For example, controlled conversion of the $1T^{\prime}-$MoTe$_2$ phase to the $2H-$phase, as recently reported \cite{park_phase_engineering}, could enable circuits composed of a single material functioning as both semiconducting channels and metallic interconnects. More precise control of the phase change might also be used to minimize the metal-semiconductor Schottky barrier by continuous evolution of the electronic band structure, in order to overcome current limits on optoelectronic performance \cite{contacts}. In fact, recent work has reported contact phase engineering by laser processing\cite{phase_patterning} and chemical modification\cite{phase_engineering}.
\begin{figure*}[htp]
\begin{center}
    \includegraphics[width=9 cm]{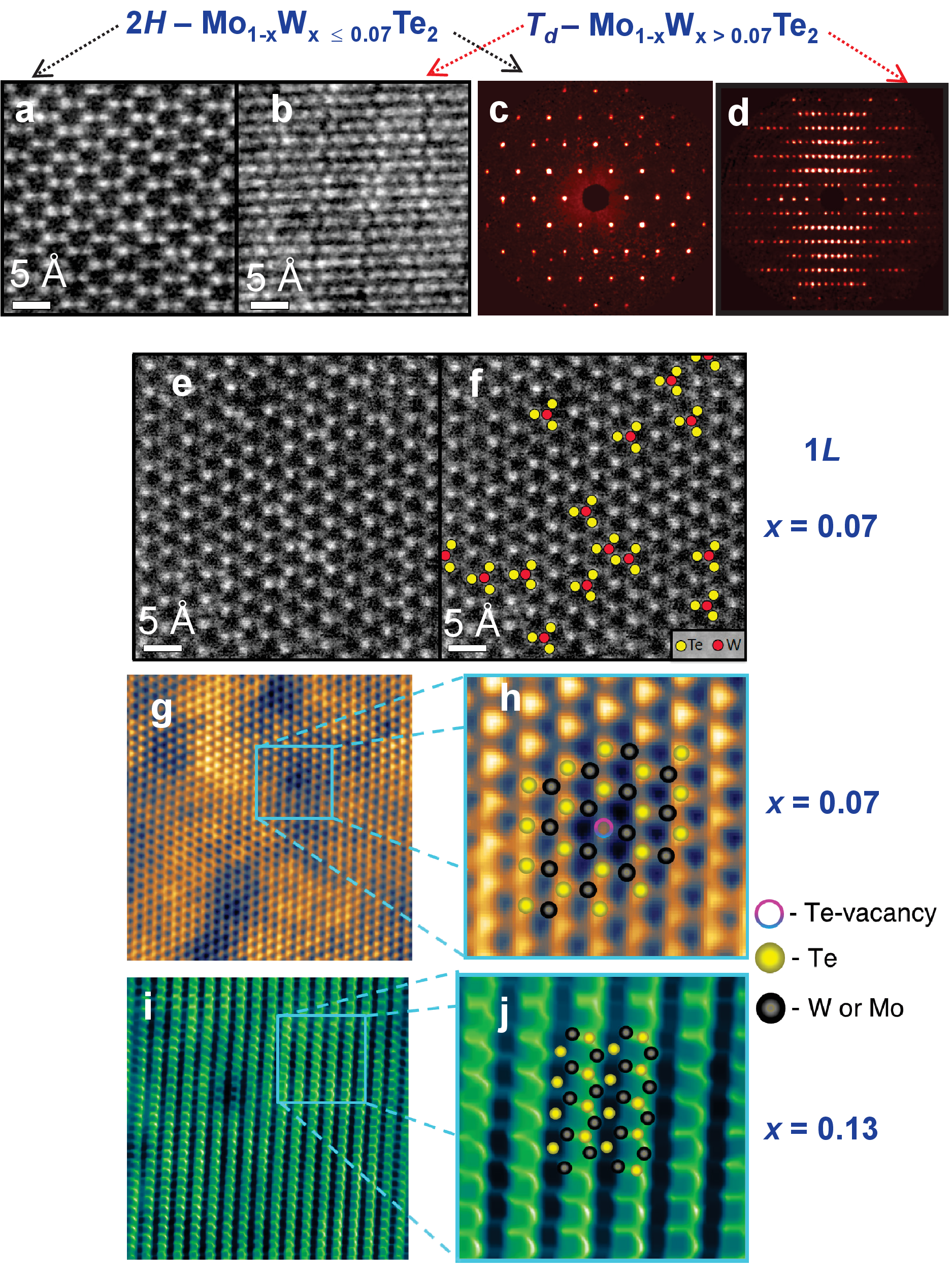}
    \caption{\textbf{$|$Structural analysis of few-layered Mo$_{1-x}$W$_x$Te$_2$ crystals.} \textbf{a}, Scanning transmission electron microscopy image of a few-layered crystal of Mo$_{1-x}$W$_x$Te$_2$ with $x \simeq 0.07$ \% displaying the hexagonal pattern typical of the $2H-$ or trigonal prismatic phase. \textbf{b}, STEM image of a few-layered crystal of Mo$_{1-x}$W$_x$Te$_2$ with $x \simeq 0.13$. Notice that its atomic arrangement is no longer consistent with the $2H-$phase. All STEM images are lightly smoothed. \textbf{c} Single-crystal \emph{X}-ray diffraction pattern for $x = 0.0$ $(00l)$ indicating that it crystallizes in the $2H-$phase. \textbf{d}, Single-crystal \emph{X}-ray diffraction pattern for $x = 0.27$ $(0kl)$ indicating that it crystallizes in the $T_d$ phase (orthorhombic $Pmn2_1$). Powder \emph{X}-ray diffraction indicates that for $x>0.08$ the Mo$_{1-x}$W$_x$Te$_2$ series crystallizes in the $T_d-$phase. \textbf{e}, $(8.25 \text{ nm})^2$ area STEM image of monolayer $2H-$Mo$_{1-x}$W$_x$Te$_2$. \textbf{f}, Brighter W atoms (indicated by red dots) are identifiable through their contrast with respect to the darker Mo atoms. Therefore, STEM indicates that these crystals are homogeneous solid-solutions containing Mo and W atoms. \textbf{g}, Scanning tunneling microscopy image of a Mo$_{1-x}$W$_x$Te$_2$ single crystal with $x = 0.07$, corresponding to an area of $(15 \text{ nm})^2$ and showing a clear hexagonal pattern as expected for the $2H-$phase. The spatial modulation in intensity reflects the local coupling between the layers. \textbf{h}, Magnification of a local area of $(2.5 \text{ nm})^2$ where one can detect a Te vacancy. \textbf{i}, STM image of a $x = 0.13$ single-crystal, also corresponding to an area of $(15 \text{ nm})^2$, showing a pattern of parallel chains as expected for the orthorhombic $T_d-$phase. \textbf{j}, Magnification of a local region of $(2.5 \text{ nm})^2$ revealing the intra-chain structure and illustrating the crystallographic positions of transition metal (black dots) and Te (yellow dots) atoms, respectively.}
\end{center}
\end{figure*}
\begin{figure*}[htp]
\begin{center}
    \includegraphics[width=11 cm]{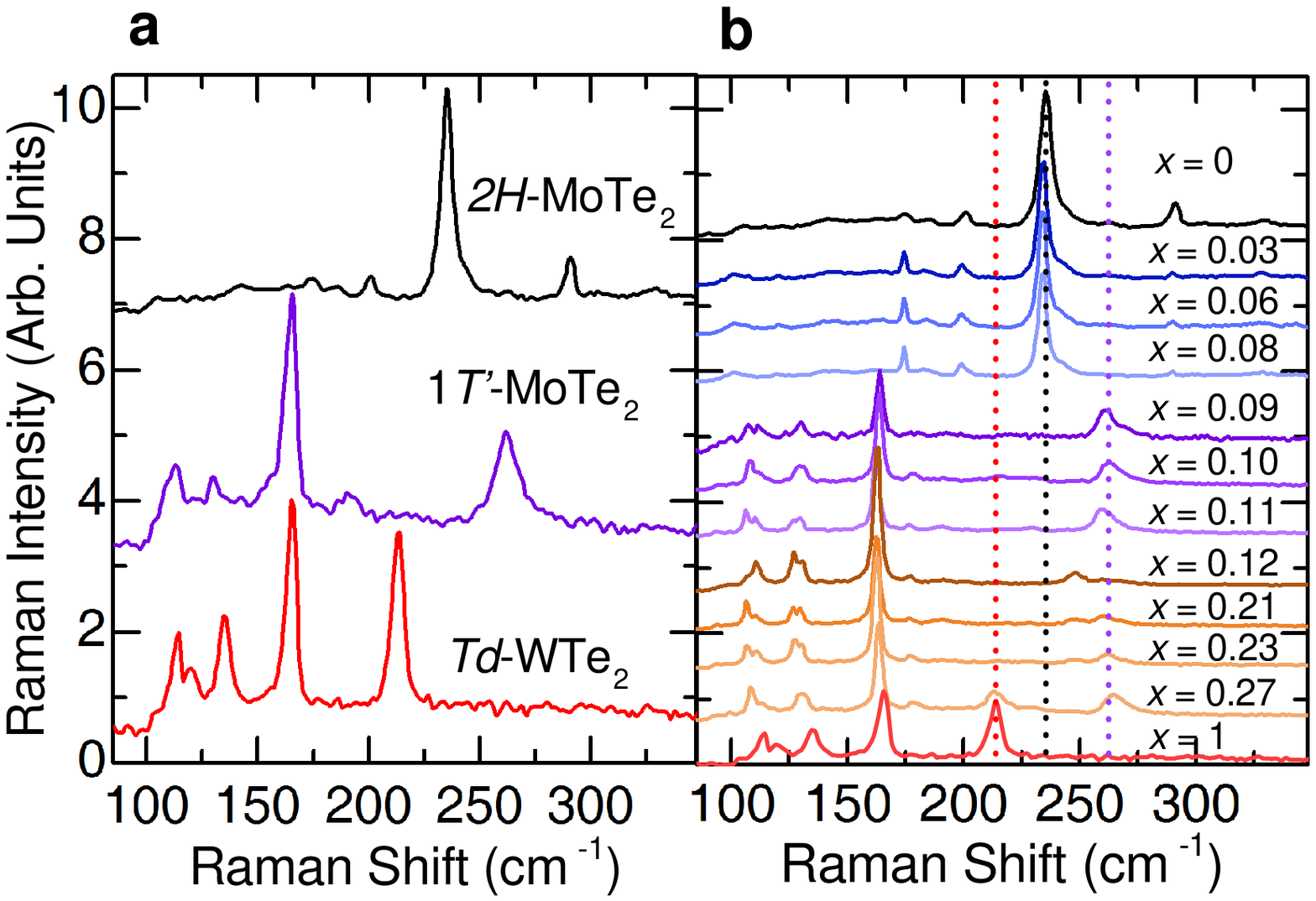}
    \caption{\textbf{$|$Raman spectra of the Mo$_{1-x}$W$_x$Te$_2$ series as a function of W doping.} \textbf{a}, Raman spectra of $2H-$MoTe$_2$ (black trace), $1T^{\prime}-$MoTe$_2$ (blue) and $Td-$WTe$_2$ (red) at room temperature using an excitation wavelength of 532 nm. \textbf{b}, Raman spectra of Mo$_{1-x}$W$_x$Te$_2$ for several values of $x$ (fraction of W). Notice the change in the spectrum observed for $x \geq 0.09$, indicating a structural phase-transition as a function of doping}.
\end{center}
\end{figure*}

The ability to phase-engineer MoTe$_2$ has many broader applications and potentially deeper implications. For instance, doping, temperature, strain, and electric fields can be used to drive metal-to-insulator transitions \cite{reed1,Alexander, Reed, longo} for sensors and nonvolatile information storage. More fundamentally, the electronic structure of mono-layers of semimetallic MoTe$_2$ (and of WTe$_2$) have been proposed to possess a $Z_2$ topological invariant characteristic of a quantum spin Hall-effect ground-state which has a gap in the bulk and non-dissipative edge states \cite{TP_transition}. If confirmed \cite{MoTe2_MI_transition}, these edge states could be used for dissipation-free nano-interconnects between logical elements based on semiconducting $2H-$MoTe$_2$ for low-power electronics. More recent theoretical developments also claim that both orthorhombic MoTe$_2$ and WTe$_2$ would be candidates for a new type of Weyl semimetallic state characterized by linear touching points between hole- and electron-Fermi surfaces, where the Berry-phase would present topological singularities \cite{bernevig,felser,bernevig2,Weyl1,Weyl2,Hasan}. These singularities, which were recently claimed to have been observed in the orthorhombic phase of MoTe$_2$ \cite{ARPES_Huang}, could lead to unconventional transport properties.

To fully control and utilize phase transitions in the two-dimensional (2D) tellurides, it is crucial to understand the phase diagram in detail. In particular, doping of the lattice can be used to precisely tune the semiconducting-metallic phase transition, and in fact W doping is known to induce a phase transition \cite{Alexander} from the $2H-$ to an orthorhombic structure, originally identified as the $1T^{\prime}-$phase, in  Mo$_{1-x}$W$_x$Te$_2$. Early studies identified a structural phase-transition from $2H$ to $T_d$ for $x > 0.15$, and with a zone of phase coexistence for $0.15 < x < 0.34$ \cite{champion}. However, given the renewed interest in this material, there is strong motivation to revisit the question of the precise evolution of the phases in the 2D tellurides with doping.


Here, we synthesize bulk crystals of Mo$_{1-x}$W$_x$Te$_2$ alloys, and characterize their composition and structure through a combination of techniques including electron microscopy, x-ray diffraction, scanning tunneling microscopy, and Raman spectroscopy. We find that W doping produces homogeneous alloys, with no phase coexistence as previously observed \cite{champion}. The structural phase transition from the semiconducting $2H-$phase towards the orthorhombic and semimetallic $T_d-$phase is sharp and occurs at a modest critical molar fraction $x_c \sim 0.08$. Since crystals with $x \lesssim x_c$ are likely to be susceptible to small perturbations such as strain or electric field, this opens the possibility of reversibly controlling the structural, and therefore electronic properties, of the Mo$_{1-x}$W$_x$Te$_2$ series. Additionally, we show through angle resolved photoemission spectroscopy that the geometry of the Fermi surface of $T_d-$Mo$_{1-x}$W$_x$Te$_2$ is remarkably similar to that of WTe$_2$, thus confirming its semimetallic character.

Single crystals of the Mo$_{1-x}$W$_x$Te$_2$ series were grown through a chemical vapor transport technique as described in Methods. Unless otherwise noted, samples were cooled slowly in order to obtain the equilibrium phase at room temperature. Their precise stoichiometry was determined through energy dispersive $X-$ray spectroscopy (EDS) and photoelectron spectroscopy (XPS), see Methods as well as Supplementary Fig. S1 for photoelectron core level spectrum of a Mo$_{1-x}$W$_x$Te$_2$ crystal and Supplementary Fig. S2 for details concerning the determination of the W content ($x \pm 0.01$). Stoichiometric MoTe$_2$ ($x=0$) and WTe$_2$ ($x=1$) were synthesized through a Te flux method. For MoTe$_2$, samples were slowly cooled to yield the $2H-$phase or quenched to room temperature to yield the metastable $1T^{\prime}-$phase.

Figure 1 shows structural analysis \emph{via} single crystal x-ray diffraction (XRD), scanning transmission electron microscopy (STEM), and scanning tunneling microscopy (STM).  For STEM, the crystals were exfoliated following a standard procedure and transferred onto a TEM grid, see Methods. Figures 1a and 1b display atomic resolution STEM images collected from two distinct multi-layered crystals with compositions of Mo$_{0.93}$W$_{0.07}$Te$_2$ and Mo$_{0.87}$W$_{0.13}$Te$_2$, respectively. These crystals display distinct crystallographic structures: $x \simeq 0.07$ shows the hexagonal pattern characteristic of the trigonal prismatic or the $2H-$ phase, while $x \simeq 0.13$ shows a striped pattern consistent with either the $1T^{\prime}-$ or the $T_d$-phase. In Supplementary Fig. S3 we have included STEM and electron diffraction images for $x \simeq 0.13$ from whose analysis we conclude
that it crystallizes in the orthorhombic $T_d$ phase. Nevertheless, in Figs. 1c and 1d we also show  single-crystal XRD patterns for $x=0.0$ and $x=0.27$, respectively. Analysis of these patterns confirms that crystals with $x \lesssim 0.07$ crystallize in the $2H-$phase, whereas crystals with $x > 0.07$, in this case $x = 0.27$, display the orthorhombic $T_d-$phase instead of the monoclinic $1T^{\prime}-$ one. Supplementary Fig. S4 shows $X-$ray diffraction patterns for Mo$_{0.91}$W$_{0.09}$Te$_2$ and Mo$_{0.82}$W$_{0.18}$Te$_2$, also indicating the $T_d-$phase for these concentrations. Our complete set of structural studies indicate that for all concentrations $x > x_c = 0.08$, W doping stabilizes the semimetallic $T_d$ phase, confirming that the structural transition is sharp and occurs at a W doping level significantly lower than previously reported \cite{champion}. Figure 1e shows a larger-scale STEM image of Mo$_{0.93}$W$_{0.07}$Te$_2$. In this image, bright dots surrounded by three additional dots (Te atoms) correspond to randomly distributed W atoms, as highlighted in Fig. 1f.  Therefore, the STEM images clearly indicate that the Mo$_{1-x}$W$_x$Te$_2$ series results from a homogeneous dilution of W atoms into a MoTe$_2$ matrix and not from the coexistence of $2H-$MoTe$_2$ and WTe$_2$ domains. This lack of phase coexistence is further confirmed by room-temperature STM imaging of vacuum-cleaved crystals, as shown in Figs. 1g and 1h. For $x=0.07$, see Fig. 1g, the equidistant distribution of Te atoms around the transition metal(s) forming an angle of $\theta = 120^{\circ}$ among them, indicates unambiguously the trigonal prismatic coordination of the $2H-$phase. In contrast, for $x=0.13$, see Figs. 1i and 1j, rows of atoms indicate a change in symmetry from triangular to (nearly) rectangular at the surface.  In amplified images, e.g. Fig. 1h, one can clearly discern Te vacancies (indicated by a multicolored dot). Therefore, we have enough resolution to observe vacancies, but we do not observe the coexistence of distinct crystallographic phases.

Thus TEM, STM, and XRD analysis yield consistent results, namely a transition from the $2H$ phase to the $T_d$ phase at  $x_c \sim 0.08$, with no phase coexistence even near the phase boundary.  These observations stand in contrast to the early work in Ref. \onlinecite{champion}, which reported a higher critical W concentration and a region of phase coexistence near the boundary. This discrepancy is likely attributable to the difference between the methods of synthesis used for each study. Having established the room temperature phase boundary between the $2H$ and the $T_d$ transition, we now turn to the temperature axis of the phase-diagram.


The structural phase transition as a function of doping is accompanied by changes in vibrational modes, as probed by Raman spectroscopy (see, Fig. 2). Figure 2a shows room-temperature Raman spectra obtained from the $2H$- and $1T^{\prime}-$MoTe$_2$ phases, and from WTe$_2$. The $2H-$ structure displays two main Raman peaks at 174 cm$^{-1}$ and 235 cm$^{-1}$ corresponding to the $A_{\text{1g}}$ and $E^1_{\text{g}}$ modes, respectively \cite{MoTe2_FETs,Yamamoto}. Reflecting its reduced symmetry, the $1T^\prime$ phase displays several peaks at lower wave-numbers. For this structure the main peaks occur at 163 cm$^{-1}$ and 260 cm$^{-1}$ and have been indexed as the $B_{\text{g}}$ and $A_{\text{g}}$ modes \cite{MoTe2_MI_transition}, respectively.  WTe$_2$ presents a spectrum having peaks occurring at 136 cm$^{-1}$ and 165 cm$^{-1}$ respectively, both of $A_{\text{g}}$ character \cite{Jiang}, in addition to a peak at $\sim$ 210 cm$^{-1}$ previously indexed as the $A^2_1$ mode. In Supplementary Fig. S6 we show Raman spectra for $T_d-$Mo$_{0.88}$W$_{0.12}$Te$_2$ as the number of layers decrease, indicating that the $T_d$-phase is stable down to the single-layer limit despite its high Mo content. Figure 2b shows Raman spectra for several stoichiometries of the Mo$_{1-x}$W$_x$Te$_2$ series. Mo$_{1-x}$W$_x$Te$_2$ crystallizes in the $2H-$phase for concentrations up to $x \sim 0.08$. For concentrations beyond this value the spectra abruptly change, as indicated by the disappearance of the $A_{\text{1g}}$ and of the $E^1_{\text{2g}}$ peaks at 174 cm$^{-1}$ and 235 cm$^{-1}$ respectively, which are observed when $x \lesssim 0.08$. These data support the conclusions reached by the structural probes above, namely a phase transition around $x_c \sim 0.08$ with no evidence for phase coexistence. Interestingly, As the W concentration increases beyond $x = 0.08$ we see the emergence of peaks which, at the first glance, would seem to be related to the $B_{\text{g}}$ and the $A_{\text{g}}$ modes of the $1T^{\prime}$-phase \cite{MoTe2_MI_transition}. However, single crystal $X-$ray diffraction shown in Fig. 1d and in Supplementary Fig. S3 clearly indicate that the Raman spectra in Fig. 2b must be associated with the orthorhombic $T_d$phase, with certain peaks shifted with respect to those of WTe$_2$ due to the high Mo content. An important observation is that Raman scattering yields nearly identical spectra for $1T^{\prime}-$MoTe$_2$ and for $T_d-$Mo$_{1-x}$W$_x$Te$_2$, for reasons that will have to be clarified through theoretical calculations. We note that this similarity might lead to misidentification of the $1T^{\prime}$ phase if Raman spectroscopy is the only method used to probe the crystal structure. In Supplementary Fig. S6, we show Raman scattering data as a function of the number of atomic layers for a crystal having $x=0.12$, which is close to the critical concentration $x_c \simeq 0.08$, indicating that it preserves its structure upon exfoliation despite its proximity to the phase-boundary. In Supplementary Fig. S7, we have include transport data, like the room temperature conductivity as a function of doping, which changes by orders of magnitude as one crosses the phase-boundary.
\begin{figure*}[htp]
\begin{center}
    \includegraphics[width=15 cm]{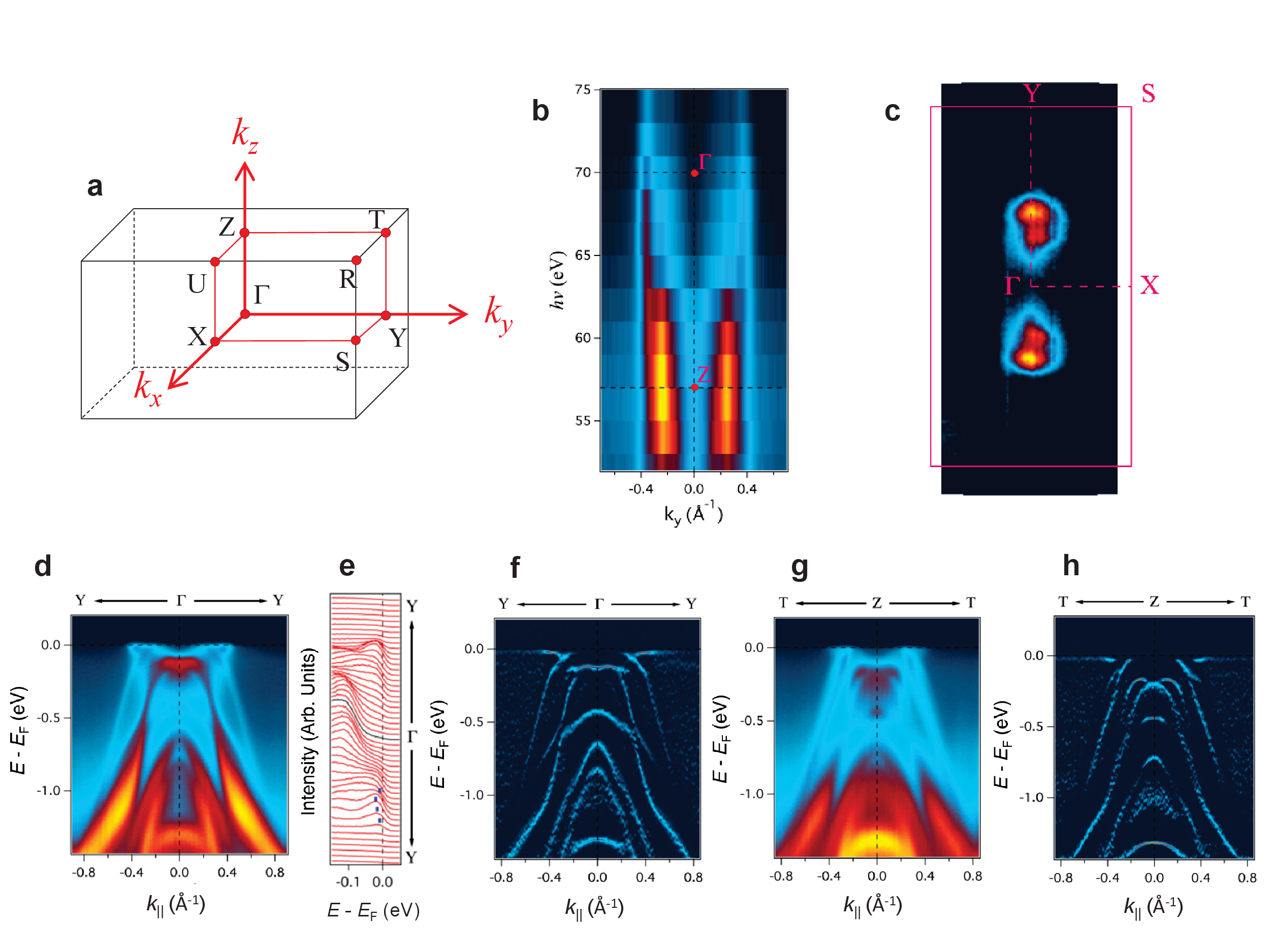}
    \caption{\textbf{$|$Angle resolved photoemission spectroscopy (ARPES) of Mo$_{1-x}$W$_x$Te$_2$.} \textbf{a}, Bulk Brillouin zone of Mo$_{0.73}$W$_{0.27}$Te$_{1.99}$ indicating its high symmetry points. \textbf{b}, ARPES intensity plot at the Fermi level as a function of both the momentum and the photon energy. \textbf{c}, Topography of the Fermi surface at $k_z = 0$. One electron- and one hole-like pocket is observed at either side of the $\Gamma-$point. The resolution of our ARPES measurements limits our ability to identify possible points of contact between the electron and the hole pockets.  \textbf{d}, ARPES band structure along Y-$\Gamma$-Y direction acquired with a photon energy of 70 eV, i.e. corresponding to $(k_z=0)$. \textbf{e}, Plot of the energy distribution curve of the low energy bands. Blue dotted line serves as a guide to the eye, indicating the positions of peaks for the electron-like band. \textbf{f}, Second derivative of the band structure collected along the Y-$\Gamma$-Y direction. \textbf{g}, ARPES band structure along T-Z-T direction acquired with a photon energy of 57 eV, i.e. corresponding to $(k_z=\pi)$. \textbf{h}, Second derivative of the band structure collected along the T-Z-T direction.}
\end{center}
\end{figure*}

Next, we investigated the electronic phase-transition accompanying the structural phase transition. In particular, while the nature of the semiconducting $2H-$phase is well understood, it is not known whether the $T_d$ phase in the W-doped material is a conventional, or a Weyl, semi-metallic system. Therefore, we investigated the electronic structure of heavily doped Mo$_{1-x}$W$_x$Te$_2$ single crystals through angle-resolved photoemission spectroscopy (ARPES), as shown in Fig. 3. The core level spectrum, shown in the Supplementary Fig. S1, displays the characteristic peaks of W and Te elements, confirming that W is alloyed into the $1T^{\prime}-$MoTe$_2$ crystal. As seen in this figure, the W $4f$ core levels have one set of doublets at 31.4 eV and 33.6 eV (right inset in Supplementary Fig. S1) respectively, in perfect agreement with the values found in the literature \cite{Arpes_book}. Meanwhile the Te $4d_{5/2}$ and $4d_{3/2}$ doublets split into four peaks (left inset in Supplementary Fig. S1). This suggests that the Mo/W layer is sandwiched by the Te layers, making the Te layer the exposed surface.
To investigate the electronic structure along the $k_z$-direction of the three-dimensional Brillouin zone (BZ), which is depicted in Fig. 3a,
we performed photon-energy-dependent ARPES measurements with energies ranging from 40 to 90 eV. Figure 3b shows the ARPES spectra at the Fermi level $E_F$ as a function of the momentum and photon energy from 55 to 75 eV. We extracted the positions of the $\Gamma$ $(k_z=0)$ and $Z$ $(k_z = \pi)$ points from the dispersion as a function of $k_z$, as shown. Figure 3c shows the Fermi surface of Mo$_{0.73}$W$_{0.27}$Te$_2$ acquired at $h\nu = 70$ eV. The Fermi surface along the Y-$\Gamma$-Y direction shows two hole-pockets and two-electron pockets at either side of $\Gamma$ which would seem to touch. This geometry for the Fermi surface of Mo$_{0.73}$W$_{0.27}$Te$_2$ (as well as its overall electronic band-structure) is remarkably similar to the one reported in Ref. \onlinecite{Pletikosic} for WTe$_2$ and therefore remarkably different from the one already reported \cite{ARPES_Huang} for orthorhombic MoTe$_2$. This difference is particularly striking given its considerably larger content of Mo relative to W. Notice that such a simple Fermi surface would be in broad agreement with our recent study \cite{QOs} on the quantum oscillatory phenomena observed in $T_d$-MoTe$_2$.
\begin{figure*}
\begin{center}
	\includegraphics[width = 10 cm]{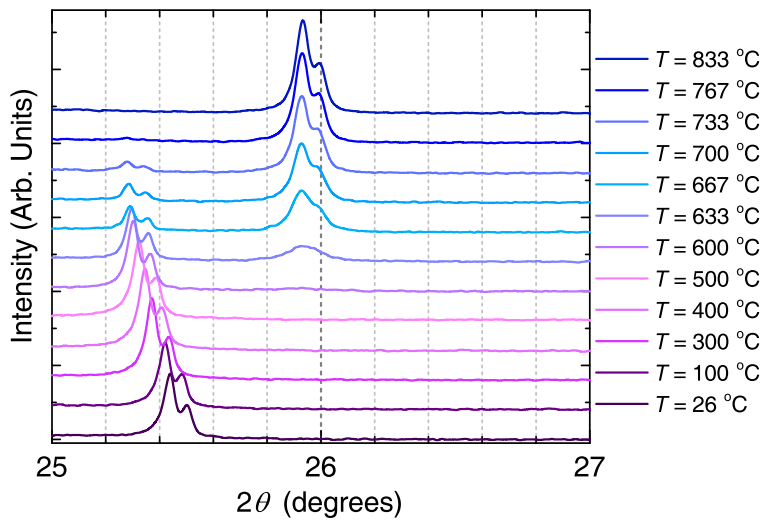}
	\caption{$|$ \textbf{Powder X-ray diffraction as a function of angle for $2H$-Mo$_{.95}$W$_{.05}$Te$_2$.}  Notice the disappearance of the peaks associated with the $2H$-phase and the emergence of new peaks, above T = 600 $^{\circ}C$ which can be ascribed to the $1T^\prime$-phase.
In a certain range of temperatures the coexistence of Bragg reflections associated to both phases results from the coexistence of domains and indicates a first-order phase-transition.}
\end{center}
\end{figure*}

ARPES band-maps along the high symmetry directions of a Mo$_{0.73}$W$_{0.27}$Te$_2$ single crystal, as well as the corresponding plots of their second derivatives, are shown in Figs. 3d through 3h. Figs. 3d and 3f show band-maps, and corresponding second derivative, acquired with a photon energy of 70 eV $(k_z=0)$ along the Y-$\Gamma$-Y direction. Figs. 3g and 3h correspond to band maps and second derivatives collected along the T-Z-T direction with a photon energy of 57 eV corresponding to $(k_z= \pi)$. The remarkable features near $E_F$ are the flat hole-like band crossing $E_F$ around $k_{\|} \sim 0.2$ \AA$^{-1}$, and an electron-like pocket in the vicinity of $k_{\|} \sim 0.4$ \AA$^{-1}$. The band connecting the hole- and the electron-like pockets is assigned to a surface state, which have already been claimed to be topologically nontrivial \cite{Hasan, Belopolski}. When compared to the calculations in Ref. \onlinecite{Hasan} the conduction band minimum is observed to be very close to the Fermi level, which makes this surface state not as easily detectable as one would expect from the calculations. The surface state is more clearly exposed in Supplementary Fig. S8. Notice that the bands near $E_F$ at $k_z=0$ have higher binding energies than those at $k_z=\pi$. As a result, the electron pocket and the surface state become more apparent in Figs. 3d and 3f.

Having established the room temperature boundary between the $2H-$ and $T_d-$ phases and explored the electronic structure of the latter phase, we now turn to the temperature axis of the phase-diagram. Figure 4 shows powder XRD patterns for a sample with $x \simeq 0.05$, at different temperatures upon heating from room-temperature. Above $T = 600$ $^\circ$C, the peaks associated with the $2H-$phase disappear and new peaks that can be ascribed to the $1T^\prime-$phase appear. Similar studies for different compositions are shown in Supplementary Fig. S5. We find that the boundary is situated at $T_{2H-1T^{\prime}} \sim 650 ^{\circ}$C with a large, sample dependent uncertainty of the order of $\sim 50 ^{\circ}$C previously attributed to variations in the Te stoichiometry \cite{MoTe2_MI_transition}. The variation of $T_{2H-1T^{\prime}}$ as a function of $x$ remains within this uncertainty, therefore the boundary should be considered as doping independent. We do not see evidence for an extended region in temperature where both phases would coexist\cite{MoTe2_MI_transition}. The $1T^{\prime}-$MoTe$_2$ phase continues to display a good degree of crystallinity at high $T$s indicating that the structural transition is not driven by an increase in the number of Te vacancies or material degradation.
\begin{figure}[htp]
\begin{center}
    \includegraphics[width=6 cm]{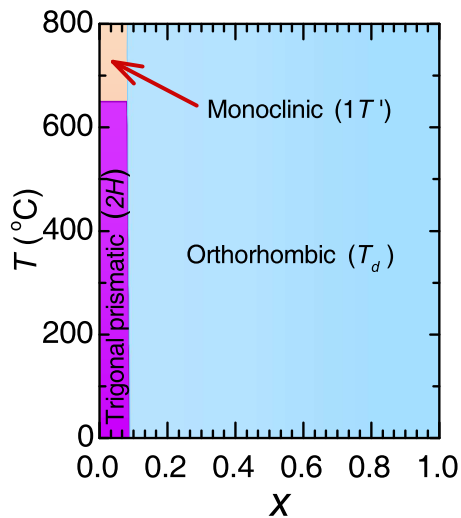}
    \caption{\textbf{$|$Resulting temperature ($T$) as a function of the doping fraction, $x$, phase-diagram of the Mo$_{1-x}$W$_x$Te$_2$ series.} Overall phase-diagram based on the array of experimental techniques used for this study. The phase-boundary between the $2H-$, $1T^{\prime}$ and the $T_d$ phases were determined through powder $X-$ray diffraction measurements as a function of $T$ and $x$ up to $x=0.13$. Above 800 $^{\circ}$C the samples decompose or lose their crystallinity. }
\end{center}
\end{figure}

The proposed phase diagram, shown in Figure 5, depicts a sharp phase-boundary between the $2H-$ and the $T_d$ phases at $x_c \sim 0.08$, and the boundary between the $2H-$ and $1T^{\prime}-$ phases at $\sim 650 ^{\circ}$C. In Supplementary Fig. S9 we compare $X-$ray powder diffraction data among samples crystallizing in the $2H-$ and in the $T_d-$phases and the role of the temperature. The important point is that, in contrast to the $2H-$phase, and even for samples with a W concentration very close to the critical one, we could not detect a structural phase-transition as a function of $T$ in samples crystallizing in the $T_d-$phase. Given that the orthorhombic $T_d-$phase becomes the ground state of $1T^{\prime}-$MoTe$_2$ and the larger area occupied by it in the phase diagram, one is led to conclude that it is thermodynamically more stable than the latter phase. The most remarkable feature of the phase diagram is the very small concentration in W required to stabilize the orthorhombic semi-metallic $T_d-$phase, and not the coexistence of the $2H-$ and the $1T^{\prime}-$ phases as predicted by Ref. \onlinecite{Alexander}, through a sharp boundary situated at $x = 0.08 \pm 0.01$. Such a sharp boundary points to a first-order phase-transition as a function of doping with the caveat that we could not detect phase coexistence.

It is quite remarkable that a semiconducting band gap as large as $\sim 1$ eV\cite{Heinz} for $2H-$MoTe$_2$, can be entirely suppressed by substituting just $x_c \simeq 9 \pm 1$ \% of Mo for W which stabilizes a semimetallic state, as clearly indicated by angle resolved photoemission experiments. Here, the situation bears a certain resemblance with the transition metal oxides such as the cuprates, whose charge- or Mott-gap is estimated to be $\Delta \sim 2$ eV, but where a small concentration of dopants, in the order of 5 \%, is enough to stabilize a metallic state (albeit anomalous) and even superconductivity \cite{Patrick}. This clearly indicates that both the structural and the concomitant electronic phases of MoTe$_2$ are particularly susceptible to small perturbations. This suggests that it should be possible to reversibly induce structural-transitions through the application of strain \cite{Alexander} or an electric field \cite{Reed}, particularly in $2H-$Mo$_{1-x}$W$_x$Te$_2$ crystals with $x \lesssim x_c$. This would make the $2H-$Mo$_{1-x}$W$_x$Te$_2$ series particularly appealing for the development of phase-change memory devices\cite{Alexander, reed1, Reed, zhang} or for a new generation of optoelectronic devices, whose metallic interconnects could be created or ``erased" at will through the application of an electrical signal, instead of a chemical treatment. Finally, the fact that the Mo$_{1-x}$W$_x$Te$_2$ series produce homogeneous alloys, is not only a major result of this study, but opens the unique possibility of exploring the evolution of their predicted topological/electronic properties \cite{Hasan}, and of perhaps detecting topological phase-transitions in the bulk as well as in the surface state through the evolution of the Fermi surface. In effect, ARPES indicates that the Fermi surfaces of $T_d-$MoTe$_2$ (Ref. \onlinecite{ARPES_Huang}) and of $T_d-$Mo$_{1-x}$W$_x$Te$_2$ (this work) are remarkably different, a fact that can only be reconciled with an electronic/topological phase-transition as a function of W doping. In effect, since W doping tends to stabilize the $T_d$-phase, it is reasonable to expect that one can stabilize it also in samples containing small amounts of W by quickly cooling the crystals to room temperature during the synthesis process. This would produce a phase-diagram not containing the region originally occupied by the $2H-$phase. This set of orthorhombic samples would allow us to explore the evolution of the Fermi surface as a function of W doping to understand, for example, how the large hole-pocket seen by ARPES at the center of the Brillouin zone \cite{ARPES_Huang} disappears to originate the hole-pockets seen by us at either side of zone center. Such electronic phase-transition should lead to either the suppression or the displacement of the Weyl-points already seen by ARPES, or to a concomitant topological phase-transition as a function of W doping.
\section*{Methods}
\subsection*{Sample synthesis} For the synthesis of pure $2H-$ and $1T'-$MoTe$_2$ as well as WTe$_2$  Mo, 99.9999\%, W, 99.9999 \% and Te 99.9999 \% were placed in a quartz ampoule in a ratio of 1:25 for growth in a Te flux. Subsequently, the material was heated up to 1050$^{\circ}$C and held there for 1 day. Then, the ampoule was slowly cooled down to 525$^{\circ}$C to yield either $T_d-$WTe$_2$ or $2H-$MoTe$_2$ and then centrifuged. To produce the $1T^{\prime}-$MoTe$_2$ phase crystals were centrifuged at 900$^{\circ}$C. The ``as harvested" single crystals were subsequently annealed for a few days at a temperature gradient to remove the excess Te. For Mo$_{1-x}$W$_x$Te$_2$, single crystals were synthesized through a chemical vapor transport technique using iodine or TeCl$_4$ as the transport agent. Samples were held at 750 $^\circ$C with a 100 $^\circ$C temperature gradient for 1 week, then subsequently cooled over 3 days to 400 $^\circ$C and removed from the furnace. Each growth commonly yielded crystals of both structure types ($T_d-$ and $2H-$), except for those crystals very rich in Mo, i.e. $x < 0.05$. Stoichiometry was determined by energy dispersive $X-$ray spectroscopy (EDS) analysis through a field-emission scanning electron microscope (FEI Nova 400). A more precise stoichiometric determination was achieved using $X$-ray photoelectron spectroscopy (XPS) either at the Shanghai Synchrotron Radiation Facility or at the Stanford Synchrotron Radiation Lightsource.
\subsection*{Scanning transmission electron microscopy}
For scanning transmission electron microscopy imaging we used a JEOL 2200FS spherical aberration corrected tool operated under 200 kV.
When using a 25.6 mrad convergence angle our probe size was 0.9 \AA.  Although 200 kV is most likely above the sample damage threshold,
we used limited acquisition times and beam exposure to minimize the possible changes to the sample structure. Micro exfoliated
few-layered samples were transferred onto TEM grids \emph{via} a dry transfer method using polypropylene carbonate.
\subsection*{X-ray diffraction as a function of the temperature }
Powder samples of Mo$_{1-x}$W$_x$Te$_2$ were prepared by sonicating chemical vapor transport grown bulk crystals in hexane.
The Mo$_{1-x}$W$_x$Te$_2$ dispersion was then drop cast onto $c$-axis sapphire substrates. This preparation led to highly textured powders,
with the $c$-axis of the sample roughly aligned with the substrate surface normal. Heating X-ray diffraction measurements were carried out
in a PANalytical X’Pert 2 diffractometer with an Anton-Paar domed hot stage that was purged with ultra pure nitrogen.
X-rays were generated from a copper target, with the Cu K$_{\beta}$ radiation removed by using a nickel filter. Several samples were also prepared via exfoliation of bulk crystals onto sapphire substrates. The phase-transition temperature was found to be independent on the method used.
\subsection*{Scanning tunneling microscopy}
Scanning tunneling microscopy (STM) measurements were performed with a home built
variable temperature, ultra high vacuum STM system at $T = 82$ K. Single
crystalline Mo$_{1-x}$W$_x$Te$_2$ was mounted onto metallic sample holders using a vacuum safe
silver paste. Samples were transferred into the STM chamber and
cleaved in-situ to expose a clean surface on which measurements were
performed. The Pt-Ir STM tip was cleaned and calibrated against a gold
(111) single crystal prior to the measurements.
\subsection*{Angle resolved photoemission spectroscopy}
ARPES measurements were performed at the Dreamline beamline of the Shanghai Synchrotron Radiation Facility with a Scienta D80 analyzer.
The energy and angular resolutions were set to 15 meV and $0.2^{\circ}$, respectively.
The ARPES data were collected using horizontally-polarized light with a vertical analyzer slit.
The samples were cleaved in situ and measured at $T=40$ K in a vacuum better than $5 \times 10^{-11}$ Torr.
The cleaved surfaces are observed to be flat at a scale $> 100$ $\mu$m while the beam spot size of the incident light is $30 \times 20$  $\mu$m$^2$, therefore the electronic structure probed by us is from a single domain.


\begin{thebibliography}{0}
\bibitem{Revolins} Revolins, E.\ \& Beerntse D. J.\  Electrical properties of $\alpha$- and $\beta$-MoTe$_2$ as affected by
stoichiometry and preparation temperature. \emph{J.\ Phys.\ Chem.\ Solids} \textbf{27}, 523-526 (1966).\
\bibitem{Vellinga} Vellinga, M. B., de Jonge, R., \& Haas, C. Semiconductor to metal transition in MoTe$_2$
\emph{J. Solid State Chem.} \textbf{2}, 299-302 (1970).\
\bibitem{review1} Voiry, D.\, Mohite, A.\, \& Chhowalla, M.\ Phase engineering of transition metal dichalcogenides. \emph{Chem. Soc. Rev.} \textbf{44}, 2702-2712 (2015).\
\bibitem{MoTe2_MI_transition} Keum, D. H., \emph{et al}.  Bandgap opening in few-layered monoclinic MoTe$_2$. \emph{Nature Phys.} \textbf{11}, 482-486 (2015).
\bibitem{Heinz} Ruppert, C.\,  Aslan, O. B.\, \& Heinz, T.\ F.\ Optical Properties and Band Gap of Single- and Few-Layer MoTe$_2$
Crystals. Nano Lett. \textbf{14}, 6231–6236 (2014).\
\bibitem{bernevig} Soluyanov, A.\ A.\ \emph{et al}. A New Type of Weyl Semimetals. \emph{Nature} \textbf{527}, 495-498 (2015).\
\bibitem{felser} Sun, Y.\, Wu, S.\-C.\, Ali, M.\ N.\, Felser, C.\, \& Yan, B.\ Prediction of the Weyl semimetal in the orthorhombic MoTe$_2$.
 \emph{Phys. Rev. B} \textbf{92}, 161107 (2015).\
\bibitem{bernevig2} Wang, Z.\ \emph{et al}. MoTe$_2$: A type-II Weyl Topological Metal. \emph{Phys. Rev. Lett.} \textbf{117}, 056805 (2016).\
\bibitem{TP_transition} Qian, X.\ F.\, Liu, J.\ W.\,  Fu, L.\, \&  Li, J.\ Quantum spin Hall effect in two-dimensional transition metal dichalcogenides. \emph{Science} \textbf{346}, 1344 (2014).\
\bibitem{reed1} Duerloo, K.\ A.\ N.\, Li, Y.\, Reed, E.\ J.\ Structural phase transitions in two-dimensional Mo- and W-dichalcogenide monolayers.
\emph{Nat. Commun.} \textbf{5}, 4214 (2014).\
\bibitem{reed2} Duerloo, K.\ A.\ N.\ \& Reed, E.\ J.\  Structural Phase Transitions by Design in Monolayer Alloys.
\emph{ACS Nano} \textbf{10}, 289-297 (2016).\
\bibitem{phase_engineering} Kappera, R. \emph{et al}. Phase-engineered low-resistance contacts for ultrathin MoS$_2$ transistors.
\emph{Nature Mater.} \textbf{13}, 1128-1134 (2014).\
\bibitem{phase_patterning} Cho, S. \emph{et al}. Phase patterning for ohmic homojunction contact in MoTe$_2$.
\emph{Science} \textbf{349},  625-628 (2015).\
\bibitem{Reed} Li, Y., Duerloo, K.-A. N., Wauson, K. \& Reed, E. J. Structural semiconductor-to-semimetal phase transition in two-dimensional materials induced by electrostatic gating. \emph{Nat. Commun.} \textbf{7}, 10671 (2016).\
\bibitem{zhang} Zhang, C.\ \emph{et al}. Charge Mediated Reversible Metal−Insulator Transition in Monolayer MoTe$_2$ and W$_x$Mo$_{1−x}$Te$_2$ Alloy. \emph{ACS Nano} \textbf{10}, 7370−7375 (2016).\
\bibitem{pletikosic}Pletikosi\'{c}, I., \ Ali, M.\ N., Fedorov, A.\ V.\, Cava, R.\ J.\,  \&  Valla, T.\ Electronic Structure Basis for the Extraordinary Magnetoresistance in WTe$_2$, \emph{Phys. Rev. Lett.} \textbf{113}, 216601 (2014).\
\bibitem{Lezama} Lezama, I.\ G.\, Ubaldini, A.\, Longobardi, M.\, Giannini, E.\, Renner, C.\, Kuzmenko, A.\ B.\, Morpurgo, A.\ F.\
 Surface transport and band gap structure of exfoliated $2H$-MoTe$_2$ crystals. \emph{2D Materials} \textbf{1}, 021002 (2014).\
\bibitem{Morpurgo_crossover} Lezama, I.\ G.\,  Arora, A.\, Ubaldini, A.\, Barreteau, C.\, Giannini, E.\, Potemski, M.\, Morpurgo, A.\ F.\
Indirect-to-Direct Band Gap Crossover in Few-Layer MoTe$_2$. \emph{Nano Lett.} \textbf{15}, 2336-2342 (2015).\
\bibitem{MoTe2_FETs} Pradhan, N. R. \emph{et al}. Field-Effect Transistors Based on Few-Layered alpha-MoTe$_2$. \emph{ACS Nano} \textbf{8},  5911-5920 (2014).\
\bibitem{MoTe2_ambipolar} Lin, Y.\ F.\ \emph{et al}. Ambipolar MoTe$_2$ Transistors and Their Applications in Logic Circuits. \emph{Adv. Mater.} \textbf{26}, 3263–3269 (2014).\
\bibitem{Mak_review} Mak, K. F. \& Shan, J. Photonics and optoelectronics of 2D semiconductor transition metal dichalcogenides. \emph{Nature Photon.} \textbf{10}, 216-226 (2016).\
\bibitem{park_phase_engineering} Park, J.\ C.\ \emph{et al}. Phase-Engineered Synthesis of Centimeter-Scale $1T^{\prime}$- and $2H$-Molybdenum Ditelluride Thin Films.
\emph{ACS Nano} \textbf{9}, 6548-6554 (2015).\
\bibitem{contacts} Zhang, W.\, Chiu, M.\ H.\, Chen, C.\ H.\, Chen, W.\, Li, L.\ J.\, \& Wee, A.\ T.\ S.\
Role of Metal Contacts in High-Performance Phototransistors Based on WSe$_2$ Monolayers. \emph{ACS Nano} \textbf{8}, 8653-8661 (2014).\
\bibitem{Alexander} Alexander, K.-, Duerloo N. \& Reed, E. J. Structural Phase Transitions by Design in
Monolayer Alloys. \emph{ACS Nano} \textbf{10}, 289-297 (2016).\
\bibitem{longo} Zhang, C., KC, S., Nie, Y., Liang, C., Vandenberghe, W. G., Longo, R. C., Zheng, Y., Kong, F., Hong, S., Wallace, R. M. \& Cho, K. Charge Mediated Reversible Metal-Insulator Transition in Monolayer MoTe$_2$ and W$_x$Mo$_{1−x}$Te$_2$ Alloy, \emph{ACS Nano} \textbf{10}, 7370-7375 (2016).\
\bibitem{Weyl1} Weng, H.\ M.\, Fang, C.\, Fang, Z.\, Bernevig, B.\ A.\, \& Dai, X.\ Weyl Semimetal Phase in Noncentrosymmetric Transition-Metal Monophosphides. \emph{Phys.\ Rev.\ X}  \textbf{5}, 011029 (2015)
\bibitem{Weyl2} Xu, S.\ Y.\ \emph{et al}. Discovery of a Weyl fermion semimetal and topological Fermi arcs. \emph{Science} \textbf{349}, 613-617 (2015).\
\bibitem{Hasan} Chang, T. R. \emph{et al}. Prediction of an arc-tunable Weyl Fermion metallic state in Mo$_x$W$_{1-x}$Te$_2$, \emph{Nat. Commun.} \textbf{7}, 10639 (2016).\
\bibitem{ARPES_Huang} Huang, L. \emph{et al}. Spectroscopic evidence for a type II Weyl semimetallic state in MoTe$_2$. \emph{Nat. Mater.} doi:10.1038/nmat4685 (2016).\
\bibitem{champion} Champion, J.\ A.\ Some propertities of (Mo,W)(Se,Te)$_2$. \emph{Brit. J. Appl. Phys.} \emph{16}, 1035 (1965).\
\bibitem{Yamamoto} Yamamoto, M. \emph{et al}. Strong Enhancement of Raman Scattering from a Bulk-Inactive Vibrational Mode in Few-Layer MoTe$_2$. \emph{ACS Nano} \textbf{8}, 3895-3903 (2014).\
\bibitem{Jiang} Jiang, Y. C. \emph{et al}. Raman fingerprint for semimetal WTe$_2$ evolving from bulk to monolayer. \emph{Sci. Rep.} \textbf{6}, 19624 (2016).\
\bibitem{Arpes_book} Chastain, J., \& King, R. C. Eds. Handbook of X-ray photoelectron spectroscopy: a reference book of standard spectra for identification and interpretation
of XPS data. Eden Prairie, MN: Physical Electronics, 1995.
\bibitem{Pletikosic} Pletikosic, I. Ali, M. N., Fedorov, A. V., Cava, R. J., \& Valla, T. Electronic Structure Basis for the Extraordinary Magnetoresistance in WTe$_2$.
\emph{Phys. Rev. Lett.} \textbf{113}, 216601 (2014).\
\bibitem{Belopolski} Belopolski, I. \emph{et al}. Fermi arc electronic structure and Chern numbers in the type-II
Weyl semimetal candidate Mo$_x$W$_{1−x}$Te$_2$. \emph{Phys. Rev. B} \textbf{94}, 085127 (2016).\
\bibitem{QOs} Rhodes, D. \emph{et al}. Impurity dependent superconductivity, Berry phase and bulk Fermi surface of the Weyl type-II semimetal candidate MoTe$_2$. \emph{arXiv:1605.09065} (2016).\
\bibitem{Patrick} Lee, P. A., Nagaosa, N. \&  Wen, X.-G. Doping a Mott Insulator: Physics of High Temperature Superconductivity. \emph{Rev. Mod. Phys.} \textbf{78}, 17-85 (2006).


\end{thebibliography}

\begin{acknowledgments}
$^{\dag}$These authors contributed equally to this work.
The subsequent order of authorship do not reflect the relative importance
among the contributions from the different authors and groups.
Their contributions to this work should be considered of equal relevance.
L.~B. is supported by the U.S. Army Research Office MURI Grant W911NF-11-1-0362.
This work was supported in part by the Molecular and Electronic Nanostructures theme of the Beckman Institute at UIUC.
Electron microscopy work was performed at the Frederick Seitz Materials Research Laboratory Central Research Facilities, University of Illinois.
Single-crystal X-ray diffraction was performed in the Shared Materials Characterization Laboratory at Columbia University.
AML acknowledges support by the U.S. Department of Energy, Basic Energy Sciences, Materials Sciences and Engineering Division.
The work of R.M.O., J.I.D., W.J., and Y.L. was financially supported by the U.S. Department of Energy under Contract No. DE-FG 02-04-ER-46157.
F.E. gratefully acknowledges Grant LPDS 2013-13 from the German National Academy of Sciences Leopoldina. This work was also supported by the DOE-BES, Materials Sciences and Engineering Division, under Contract DE-AC02-76SF00515 and by the W. M. Keck Foundation and the Gordon and Betty  Moore Foundation’s EPiQS Initiative through Grant No. GBMF4545.
D.C., N.F., A.A. and J.H. acknowledge support from AFOSR grant FA9550-14-1-0268.
R.L. and S.C.W. were supported by the National Natural Science Foundation of China (No. 11274381).
J.Z.M., T.Q., and H.D. were supported by the Ministry of Science and Technology of China (No. 2015CB921300, No. 2013CB921700),
the National Natural Science Foundation of China (No. 11474340, No. 11234014), and the Chinese Academy of Sciences (No. XDB07000000).
STM work is supported by AFOSR (FA9550-11-1-0010, DE) and NSF (DMR-1610110, ANP).
This research used resources of  (XPS at) the Center for Functional Nanomaterials, which is a U.S. DOE Office of Science Facility,
at Brookhaven National Laboratory under Contract No. DE-SC0012704.
The NHMFL is supported by NSF through NSF-DMR-1157490 and the State of Florida.
\end{acknowledgments}

\section*{Author contributions}
D.R. synthesized and characterized the single crystals through electron dispersive spectroscopy, Raman scattering and transport measurements.
Y.-c.C., W.Z., and Q.R.Z. were directly involved in the synthesis and in the preliminary characterization of the crystals.
D.C., T.K., A.M., M.C., M.D., N.F., A.A., and J.H. motivated the project, performed Raman, photoluminescense and
device characterization. B.E.J. and P.Y.H. performed atomic resolution STEM transmission electron microscopy.
W.J., J.Z.M., R.L., T.S., S.C.W., T.Q., H.D., J.I.D., and R.M.O. performed angle-resolved photoemission spectroscopy measurements.
Y.L., X.T., T.S., J.I.D., and R.M.O. performed x-ray photoemission spectroscopy measurements.
D.E. and A.N.P. performed scanning tunneling spectroscopy measurements. 
C.N., E.M., F.E., and A.M.L. performed powder X-ray and electron diffraction measurements as function of the temperature.
D.W.P. Performed single crystal $X-$ray diffraction.
D. R., J. H. and L. B. wrote the manuscript with the input of all co-authors.

\section*{Additional information}
Supplementary information is available in the online version of the paper. Reprints and
permissions information is available online at www.nature.com/reprints.
Correspondence and requests for materials should be addressed to L.B.
\section*{Competing financial interests}
The authors declare no competing financial interests.

\end{document}